\documentclass[aps,prx,floatfix,amsmath,amssymb,twocolumn]{revtex4-1}%preprint
\usepackage{amsmath, amssymb, graphics, mathrsfs, bm, stackrel,bbold}
\usepackage{subfigure}%http://texblog.wordpress.com/2007/08/28/placing-figurestables-side-by-side-subfigure/
\usepackage[usenames,dvipsnames]{xcolor}%http://en.wikibooks.org/wiki/LaTeX/Colors
\definecolor{Highlight}{rgb}{1,1,0.75}

\allowdisplaybreaks[1]
\usepackage{graphicx}
\usepackage[bookmarks={false}, pdfauthor={Rudro Rana Biswas}, pdftitle={Title}]{hyperref}%pdftex?
\hypersetup{colorlinks=true, linkcolor=BrickRed, citecolor=Violet, filecolor=OliveGreen, urlcolor=RoyalBlue, filebordercolor={.8 .8 1}, urlbordercolor={.8 .8 0}}%http://en.wikibooks.org/wiki/LaTeX/Hyperlinks
\usepackage[all]{hypcap}
\usepackage[export]{adjustbox}%extends includegraphics options, etc

\newcommand\ba{\begin{array}}
\newcommand\ea{\end{array}}
\newcommand\nn{\nonumber}
\renewcommand\ol{\overline}
\newcommand\ri{\right}
\renewcommand\le{\left}

\newcommand{\feyn}[1]{#1\kern-0.45em/}

\renewcommand\a{\alpha}%metric symbol

\newcommand\mbA{\mbs{A}}
\newcommand\mbB{\mbs{B}}
\renewcommand\c{\psi}%accent

\renewcommand\d{\delta}%accent

\newcommand\Dd{\Delta}

\newcommand\e{\epsilon}

\newcommand\mbE{\mbs{E}}
\newcommand\f{\phi}

\newcommand\mbj{\mbs{j}}

\renewcommand\k{\kappa}%accent

\renewcommand\l{\lambda}%non-ascii letter
\renewcommand\L{\Lambda}%non-ascii letter

\newcommand\p{\pi}
\newcommand\mbp{\mbs{p}}
\newcommand\mbpp{\mbs{\p}}
\newcommand\mbP{\mbs{P}}

\newcommand\rr{\rho}%angstrom requirement

\newcommand\mbr{\mbs{r}}
\newcommand\mbR{\mbs{R}}

\renewcommand\th{\theta}%latin char

\newcommand\mbv{\mbs{v}}
\newcommand\w{\omega}

\newcommand\vx{\chi}
\newcommand\mbx{\mbs{x}}

\newcommand\mbX{\mbs{X}}
\newcommand\y{\eta}
\newcommand\mby{\mbs{y}}

\newcommand\mbz{\mbs{z}}

\newcommand\grad{\mbs{\nabla}}
\newcommand\la{\langle}
\newcommand\ra{\rangle}
\newcommand\pd{\partial}
\newcommand\mc{\mathcal}

\newcommand\mbs{\boldsymbol}

\begin{document}
\title{Geometric response of quantum Hall states to electric fields}%Gauge invariant phase space technique in quantum Hall physics
\author{YingKang Chen}
\affiliation{Department of Physics and Astronomy, Purdue University, West Lafayette, IN 47907.}
\author{Guodong Jiang}
\affiliation{Department of Physics and Astronomy, Purdue University, West Lafayette, IN 47907.}
\author{Rudro R. Biswas}
\email{rrbiswas@purdue.edu}
\affiliation{Department of Physics and Astronomy, Purdue University, West Lafayette, IN 47907.}

%\date{\today}
\begin{abstract}
Exploiting novel aspects of the quantum geometry of charged particles in a magnetic field via gauge-invariant variables, we provide tangible connections between the response of quantum Hall fluids to non-uniform electric fields and the characteristic geometry of electronic motion in the presence of magnetic and electric fields. The geometric picture we provide motivates the following conjecture: non-uniform electric fields mimic the presence of spatial curvature. Consequently, the gravitational coupling constant also appears in the charge response to non-uniform electric fields.
\end{abstract}
%\pacs{00.00xx,yy}
\maketitle

%\tableofcontents

\section{Introduction}

Quantum Hall states were the first examples of topological quantum states of matter to be discovered \cite{1980-klitzing-uv,1982-tsui-jq}. They continue to serve as paradigmatic models of topological quantum states of matter \cite{2008-day-lp,2010-qi-ak}. The defining characteristic of these states is the topologically protected quantization of their eponymous conductance property. Less well known is the quantization of their gravitational coupling constant, which characterizes charge response to spatial curvature in the continuum \cite{1992-wen-fk} and on lattices \cite{2016-biswas-kx}. There is an active quest to understand the topological protection of the gravitational coupling constant in terms of fundamental physical principles \cite{2015-klevtsov-qr}. A flurry of recent developments have connected the gravitational coupling constant to response coefficients for a fundamentally different type of perturbation due to non-uniform electric fields. Specifically, using effective topological field theory techniques, the gravitational coupling constant is proportional to the anomalous viscosity \cite{2009-read-fk,1995-avron-fk} and appears in the current response to non-uniform electric fields  \cite{2012-hoyos-fk}. What foundational basis underlies the appearance of the gravitational coupling constant in these disparate contexts? Herein we address this question and provide tangible connections between the response of quantum Hall fluids to non-uniform electric fields, and the characteristic geometry of electronic motion in the presence of magnetic and electric fields. The geometric picture we provide motivates the following conjecture: non-uniform electric fields mimic the presence of spatial curvature. Consequently, the gravitational coupling constant also appears in the charge response to non-uniform electric fields.

%%%%%%%%%%%%%%%%%%%%%%begin%%%%%%%%%%%%%%%%%%%%%%
\begin{figure}[t]
\begin{center}
\resizebox{0.5\textwidth}{!}{\includegraphics{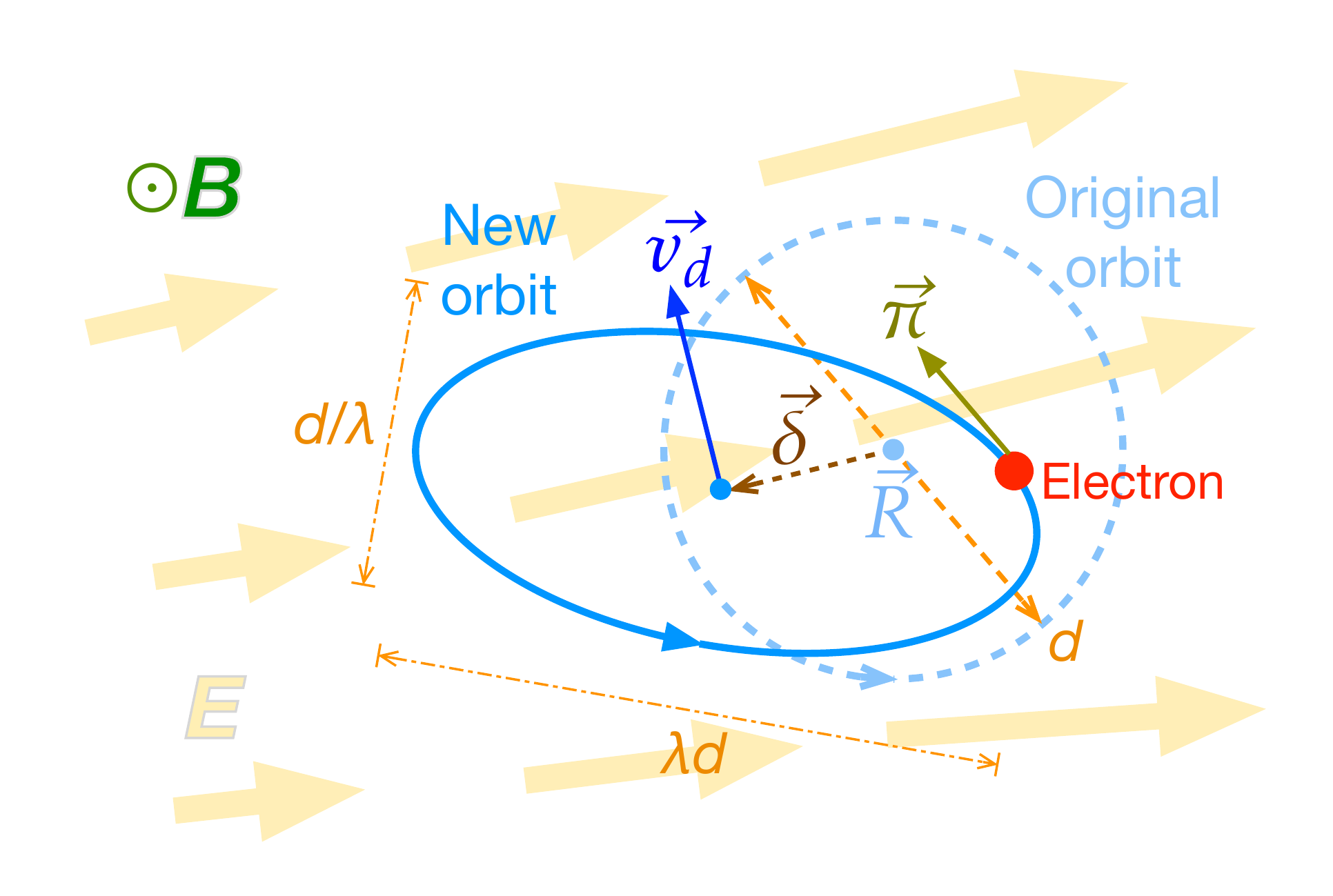}}%inclgphcs[trim=lcm bcm rcm tcm, clip=true, angle=-90]
\caption{A geometric summary of how a non-uniform electric field, $\mbE$, deforms a cyclotron orbit. The changes can be expressed in terms of a vector field, $\mbs{\Dd}(\mbR)$ (Eq.~\eqref{eq-inducedshift}), and a shearing field, $\L(\mbR)$ (Eq.~\eqref{eq-shearingmatrix}). The new orbit is shifted by amount $\mbs{\d}(\mbR) = \hat{\mbz} \times \mbs{\Dd}(\mbR)$ with respect to the original center, $\mbR$, and acquires a drift velocity, $\mbv_{d} = \mbs{\Dd}(\mbR)$. The orbit is also sheared into an ellipse with aspect ratio $\l^{2}$; see Eq.~\eqref{eq-orbitellipticity} and accompanying text for details. The guiding center coordinate, $\mbR$, labels the field-free orbit center while the kinetic momentum, $\mbpp$, gives the velocity of the electron (Eq.~\eqref{eq-GIV}).}
\label{fig-cyclotronorbit}
\end{center}
\end{figure}
%%%%%%%%%%%%%%%%%%%%%%%end%%%%%%%%%%%%%%%%%%%%%%

To elucidate the characteristic geometry of electronic motion in a magnetic field we introduce an explicitly gauge-invariant quantum description of quantum Hall systems, and thus a calculational framework naturally suited for describing attendant physics. Our formulation is based on the semiclassical description of quantum Hall physics in terms of gauge invariant variables (operators), the kinetic momenta ($\mbpp$) and guiding center coordinates ($\mbR$), respectively defined as \cite{1949-johnson-nr}
%%%%%%%%%%%%%%%%%%%%%%begin%%%%%%%%%%%%%%%%%%%%%%
\begin{align}\label{eq-GIV}
\mbpp = \mbp + e \mbA, \quad \mbR = \mbr + \le(\ell^{2}/\hbar\ri)\hat{\mbz}\times\mbpp.
\end{align}
%%%%%%%%%%%%%%%%%%%%%%%end%%%%%%%%%%%%%%%%%%%%%%
$\ell = \sqrt{\hbar/eB}$ is the magnetic length and we will orient our $z$-axis along the magnetic field, $\mbB = B \hat{\mbz}$. These operators satisfy the commutation relations:
%%%%%%%%%%%%%%%%%%%%%%begin%%%%%%%%%%%%%%%%%%%%%%
\begin{align}\label{eq-GIVcommutation0}
[R_{x}, R_{y}] = i \ell^{2}, \quad [\p_{y}, \p_{x}] = i \hbar^{2}/\ell^{2};
\end{align}
%%%%%%%%%%%%%%%%%%%%%%%end%%%%%%%%%%%%%%%%%%%%%%
$\mbpp$ and $\mbR$ commute with each other. Our quantum formalism uses these gauge-invariant operators (variables), henceforth referred to as GIVs. These replace real-space position operators and gauge-sensitive canonical momentum operators in the traditional description of the quantum Hilbert space in the Schr\"{o}dinger formalism. Here we focus on the single particle physics of electrons in the presence of a magnetic field; the extension of this formalism to the interacting case has been presented elsewhere \cite{2018-chen-zl}.

Below we show how the GIV formalism naturally and elegantly yields Landau quantization and the extensive degeneracy of Landau levels \cite{2002-yoshioka-zl,1989-zak-my,1997-zak-ug}. The geometry encapsulated by the commutation relations of the guiding center operators, Eq.~\eqref{eq-GIVcommutation0}, is reflected in the characteristic form of equations of motion of these operators in an electric field. These can be used to derive quantization of the Hall conductance in the presence of potential (electrostatic) disorder. The guiding center operators have also been used to construct the guiding center density operators \cite{1990-duncan-le} which describe collective excitations in the fractional quantum Hall phases \cite{1985-girvin-fk,1986-girvin-kx}. Recently, these operators have been used to generate a family of variational deformed Laughlin states; these were then used to obtain a geometric theory of the collective mode excitation and its connection to the anomalous non-dissipative viscosity of quantum Hall states \cite{2011-haldane-fk}.

While a plethora of works have exploited the operator algebra of the GIVs, calculations utilizing Schr\"odinger wave mechanics in the GIV Hilbert space representation are uncommon \cite{1976-boon-og,2013-biswas-wq}. In this paper we present such a calculation for non-interacting electrons in a Landau level using wavefunctions in the GIV basis to conceptualize the shearing of cyclotron orbits in the presence of non-uniform electric fields (Figure~\ref{fig-cyclotronorbit}). We find that the Hall viscosity contribution in the current response \cite{2012-hoyos-fk,2009-read-fk} is a direct consequence of shearing of cyclotron orbits. Our formalism directly connects the shearing of cyclotron orbits, i.e., a change in the effective Galilean metric, to the non-uniformity of electric fields. We calculate the effective spatial curvature induced by the electric field, Eq.~\eqref{eq-inducedcurvature} and predict that the gravitational response of integer quantum Hall states also appears in the charge response to non-uniform electric fields. We confirm our predictions with numerical calculations and conclude with a conjecture regarding the extension of these results to fractional quantum Hall states.

This paper is structured as follows: In Section~\ref{sec-GIVformalism} we define the GIV representation of the Hilbert space of planar charged particles in the presence of a uniform magnetic flux. Next, we establish how conventional wavefunctions in a variety of gauge choices can be recast in our gauge-invariant language. In Section~\ref{sec-GIVkinematics} the action of a non-uniform electric field on the cyclotron orbits is considered. In Section~\ref{sec-LLprojection} we introduce the Landau-level projection, i.e., the simplest approximation, to derive a low energy effective Hamiltonian and some basic properties of quantum Hall states. We improve this picture in Section~\ref{sec-beyondLLprojection} to derive an effective Hamiltonian which yields our main results for the geometric changes to the cyclotron orbits depicted in Figure~\ref{fig-cyclotronorbit}. These results are utilized in Section~\ref{sec-observablecalculation} to calculate, using Wigner pseudoprobability functions, the non-uniform current and charge densities induced by the electric field.

\section{Wavefunctions in the GIV formalism}\label{sec-GIVformalism}

For brevity, we set the magnetic length ($\ell$), electronic charge ($e$) and $\hbar$ to unity. In places where we consider a specific Hamiltonian we assume a quadratic dispersion for electrons with unit mass ($m$) and ignore the effects of spin.

The commutation relations between the GIVs \cite{1949-johnson-nr},
%%%%%%%%%%%%%%%%%%%%%%begin%%%%%%%%%%%%%%%%%%%%%%
\begin{align}\label{eq-GIVcommutation}
[R_{x}, R_{y}] = i, \quad [\p_{y}, \p_{x}] = i, \quad [\p_{i}, R_{j}] = 0,
\end{align}
%%%%%%%%%%%%%%%%%%%%%%%end%%%%%%%%%%%%%%%%%%%%%%
are canonical and analogous to the canonical commutation relations between the $2D$ coordinates and canonical momenta:
%%%%%%%%%%%%%%%%%%%%%%begin%%%%%%%%%%%%%%%%%%%%%%
\begin{align}%\label{eq-}
[x, p_{x}] = i , \quad [y, p_{y}] = i,\quad [(x, p_{x}), (y, p_{y})] = 0.
\end{align}
%%%%%%%%%%%%%%%%%%%%%%%end%%%%%%%%%%%%%%%%%%%%%%
Thus, the GIVs can be obtained from the canonical coordinates and momenta via a canonical transformation: a unitary transformation exists from the orthonormal quantum Hilbert space basis labeled by the coordinates, $\le\{\le|x,y\ri\ra\ri\}$, to another labeled by the values of one operator from each of the canonical pairs in Eq.~\eqref{eq-GIVcommutation}. For example, $\le\{\le|R_{x}, \p_{y}\ri\ra\ri\}$, labeled by the eigenvalues of the operators $\le(R_{x}, \p_{y}\ri)$, form one such orthonormal basis. It is well known that the Hilbert space can be labeled by any of the following orthonormal bases: $\le\{\le|p_{x},y\ri\ra\ri\}$, $\le\{\le|x,p_{y}\ri\ra\ri\}$, or $\le\{\le|p_{x},p_{y}\ri\ra\ri\}$. Analogously, alternate GIV representations are possible: $\le\{\le|R_{y}, \p_{y}\ri\ra\ri\}$, $\le\{\le|R_{x}, \p_{x}\ri\ra\ri\}$ or $\le\{\le|R_{y}, \p_{x}\ri\ra\ri\}$.

To illustrate use of this formalism, we derive the unitary transformation matrix elements $\le\la x, y | R_{x}, \p_{y} \ri\ra \equiv \vx(x,y)$, i.e., the wave function of a GIV basis state in the conventional coordinate (Schr\"odinger) representation. By definition, $\vx(x,y)$ is the simultaneous eigenstate of $\hat{R}_{x}$ and $\hat{\p}_{y}$, with eigenvalues $R_{x}$ and $\p_{y}$ respectively. Since the coordinate representation is gauge-dependent, we will need to choose a gauge for the magnetic vector potential, $\mbA$. First, we consider the Landau gauge, $\mbA_{\text{Lan}} = x \hat{\mby}$ (note that $B=1$ in our units). The eigenvalue conditions become:
%%%%%%%%%%%%%%%%%%%%%%begin%%%%%%%%%%%%%%%%%%%%%%
\begin{align}%\label{eq-}
i\pd_{y}\vx = R_{x}\vx, \quad -i\pd_{y}\vx + x \vx = \p_{y} \vx.
\end{align}
%%%%%%%%%%%%%%%%%%%%%%%end%%%%%%%%%%%%%%%%%%%%%%
The (unnormalized) solution is
%%%%%%%%%%%%%%%%%%%%%%begin%%%%%%%%%%%%%%%%%%%%%%
\begin{align}\label{eq-Landaugaugetransform}
\vx_{\text{Lan}}(x,y) \propto \d(x - R_{x} - \p_{y}) e^{-i R_{x}y}.
\end{align}
%%%%%%%%%%%%%%%%%%%%%%%end%%%%%%%%%%%%%%%%%%%%%%
Alternately, we can find $\vx(x,y)$ in the symmetric gauge, $\mbA_{\text{sym}} = \le(-y \hat{\mbx} + x \hat{\mby}\ri)/2$:
%%%%%%%%%%%%%%%%%%%%%%begin%%%%%%%%%%%%%%%%%%%%%%
\begin{align}\label{eq-Symgaugetransform}
\vx_{\text{sym}}(x,y) \propto \d(x - R_{x} - \p_{y}) e^{-i R_{x}y}e^{i \frac{x y}{2}}.
\end{align}
%%%%%%%%%%%%%%%%%%%%%%%end%%%%%%%%%%%%%%%%%%%%%%

Despite the necessity to gauge-fix $\le\{\le|x,y\ri\ra\ri\}$, the GIV states $\le\{\le|R_{x}, \p_{y}\ri\ra\ri\}$ are themselves invariant under gauge transformations. Consequently, $\vx_{\text{Lan}}$ and $\vx_{\text{sym}}$ differ only by a GIV-independent phase factor; the corresponding phase, $\f = x y/ 2$, satisfies $ \grad \f = \mbA_{\text{Lan}} - \mbA_{\text{sym}}$.

Within the GIV formalism, now consider the energy eigenstates of an electron in $2D$ experiencing a perpendicular magnetic field. Assuming minimal coupling to the gauge field, the Hamiltonian is of the form:
%%%%%%%%%%%%%%%%%%%%%%begin%%%%%%%%%%%%%%%%%%%%%%
\begin{align}\label{eq-masterHamil}
\mc{H} = K(\mbp + e \mbA) - V(\mbr) \equiv K(\mbpp) - V(\mbR + \mbpp\times\hat{\mbz}).
\end{align}
%%%%%%%%%%%%%%%%%%%%%%%end%%%%%%%%%%%%%%%%%%%%%%
Here, $K$ denotes the kinetic operator. $V$ is the local electrostatic potential which includes contributions both from externally applied fields and local electrostatic irregularities in the material; the negative sign results from the sign of electronic charge. When $V=0$, since the Hamiltonian is only a function of the kinetic momenta, the eigenstates in the GIV basis $\le\{\le|R_{x}, \p_{y}\ri\ra\ri\}$ can be found by separation of variables:
%%%%%%%%%%%%%%%%%%%%%%begin%%%%%%%%%%%%%%%%%%%%%%
\begin{align}%\label{eq-}
\Psi(R_{x}, \p_{y}) = \c(R_{x})\y(\p_{y}).
\end{align}
%%%%%%%%%%%%%%%%%%%%%%%end%%%%%%%%%%%%%%%%%%%%%%
For quadratic dispersion, $K(\mbpp) = \mbpp^{2}/2$, the commutation relations, Eq.~\eqref{eq-GIVcommutation}, imply that the Hamiltonian is equivalent to that of a quantum simple harmonic oscillator with energies \footnote{\url{http://dlmf.nist.gov/18.39.E4}} and eigenfunctions \footnote{\url{http://dlmf.nist.gov/18.39.E5}} of the form:
%%%%%%%%%%%%%%%%%%%%%%begin%%%%%%%%%%%%%%%%%%%%%%
\begin{align}\label{eq-noFieldSoln}
K(\mbpp)\eta_{n}(\p_{y}) &= \vx_{n} \eta_{n}(\p_{y}), \quad n=0,1,\ldots,\nn\\
\vx_{n} = n + \frac{1}{2},& \quad\eta_{n}(\p_{y}) = \frac{e^{-\p_{y}^{2}}H_{n}(\p_{y})}{\sqrt{2^{n}n!\sqrt{\p}}}.
\end{align}
%%%%%%%%%%%%%%%%%%%%%%%end%%%%%%%%%%%%%%%%%%%%%%
Thus, the discrete nature of the eigenvalue spectrum of $K(\mbpp)$ arises from the commutation relations satisfied by the kinetic momenta GIVs and defines the familiar electronic Landau levels (LLs). The kinetic operator does not affect the guiding center part of the wavefunction, $\c(R_{x})$, an arbitrary function, resulting in the high degeneracy of the Landau levels. This degeneracy is countable; the countability arises from the canonical commutation relation satisfied by the guiding center variables. This degeneracy can be counted using von Neumann's result that the phase space density of quantum states is $(2\p)^{-1}$ \cite{1932-von-neumann-ix}. Thus, in the guiding center phase space, i.e., the $2D$ space spanned by $(R_{x}, R_{y})$, there is one quantum state corresponding to each area of $2\p$. This area corresponds to an equal area in real space, since the guiding center corresponds to the real-space location of the cyclotron orbit center. Putting together these results and restoring units, we arrive at the well-known result: inside a Landau level, there is an extensive degeneracy arising from the existence of one quantum state per real space area of $2\p\ell^{2} = (h/e)/B$, i.e., the area pierced by one flux quantum, $\f_{0} = h/e$.

The gauge-invariant nature of the GIV quantum basis allows for straightforward visualization and representation of Landau level wavefunctions. The popularly-used wavefunctions in the Landau ($L$) and symmetric ($S$) gauges in the LL with index $n$ become simply \cite{2002-yoshioka-zl}:
%%%%%%%%%%%%%%%%%%%%%%begin%%%%%%%%%%%%%%%%%%%%%%
\begin{align}%\label{eq-}
\Psi_{L} &= \d(R_{x}-X)\eta_{n}(\p_{y}),\\
\Psi_{S} &= \eta_{m}(R_{x})\eta_{n}(\p_{y}).
\end{align}
%%%%%%%%%%%%%%%%%%%%%%%end%%%%%%%%%%%%%%%%%%%%%%
These are respectively parametrized by $X$, the $x$-coordinate of the guiding center, and $m = 0, 1, \ldots$, the eigenvalue of the operator $(R_{x}^{2}+R_{y}^{2} - 1)/2$. Note that the kinetic momentum-dependent part of the wavefunction is the same for both: it is fixed for a given LL. Thus written, these wavefunctions are gauge-independent. The corresponding Schr\"odinger wavefunctions in \emph{any} gauge can be calculated using the appropriate unitary transformation. It is straightforward to convert $\Psi_{S}$, the conventional wavefunctions used in the symmetric gauge, to the Landau gauge using the unitary transform derived in Eq.~\eqref{eq-Landaugaugetransform}:
%%%%%%%%%%%%%%%%%%%%%%begin%%%%%%%%%%%%%%%%%%%%%%
\begin{align}%\label{eq-}
&\Psi_{S}(x,y) = \iint dR_{x}d\p_{y} \; \vx(x,y) \eta_{m}(R_{x})\eta_{n}(\p_{y})\nn\\
&\propto e^{- \frac{x^{2}+y^{2}-2i xy}{4}}(x-iy)^{n-m} L_{m}^{n-m}\le(\frac{x^{2}+y^{2}}{2}\ri).
\end{align}
%%%%%%%%%%%%%%%%%%%%%%%end%%%%%%%%%%%%%%%%%%%%%%
Here, $L_{n}^{\a}$ are the associated Legendre polynomials \footnote{\url{http://dlmf.nist.gov/18.5.E12}}. This result can be verified by straightforward computation of the integral. Comparison with conventional wavefunctions in the circular gauge \cite{2002-yoshioka-zl} shows that the above expression differs only by a factor of $e^{i xy/2}$. This is as expected since the gradient of $xy/2$ accounts for the difference between the symmetric and Landau gauge vector potentials.

\section{Motion in a non-uniform electric field}\label{sec-GIVkinematics}

Consider the general Hamiltonian in Eq.~\eqref{eq-masterHamil}, with quadratically dispersing kinetic energy, which describes electronic motion in the presence of crossed uniform magnetic and \emph{non-uniform} electric fields:
%%%%%%%%%%%%%%%%%%%%%%begin%%%%%%%%%%%%%%%%%%%%%%
\begin{align}\label{eq-masterQuadHamil}
\mc{H} = \frac{(\mbp + \mbA)^{2}}{2} - V(\mbr) \equiv \frac{\mbpp^{2}}{2} - V(\mbR + \mbpp\times\hat{\mbz}).
\end{align}
%%%%%%%%%%%%%%%%%%%%%%%end%%%%%%%%%%%%%%%%%%%%%%
Our objective in this section is to calculate the local charge and current density operators as linear gradient expansions in the electrostatic potential $V$. We will also relate a subset of these expansion coefficients to apparently unrelated topological quantities, which describe electronic response to geometrical real space (gravitational) perturbations to cyclotron motion \cite{1992-wen-fk,2012-hoyos-fk}. The local charge and current operators are expressed in terms of the GIVs as follows. (Operators have been distinguished from $c$-numbers using carets in what follows. However, $\hat{\mbz}$ is the unit vector along the $z$-direction.)
%%%%%%%%%%%%%%%%%%%%%%begin%%%%%%%%%%%%%%%%%%%%%%
\begin{subequations}\label{eq-localopsinGIV}
\begin{align}
\hat{\rr}(\mbr) &= \d(\hat{\mbr} - \mbr) = \d(\hat{\mbR} + \hat{\mbpp}\times\hat{\mbz} - \mbr),\\
\hat{\mbj}(\mbr) &= \frac{\le\{\hat{\mbv},\d(\hat{\mbr} - \mbr)\ri\}}{2} = \frac{\le\{\mbpp,\d(\hat{\mbR} + \hat{\mbpp}\times\hat{\mbz} - \mbr)\ri\}}{2}.
\end{align}
\end{subequations}
%%%%%%%%%%%%%%%%%%%%%%%end%%%%%%%%%%%%%%%%%%%%%%
Here $\hat{\mbv} = i \le[\mc{H}, \mbr\ri] = \hat{\mbpp}$ is the velocity operator.

We assume that the potential $V$ is weak and varies slowly in space; specifically, that its variation over a magnetic length is negligible compared to the inter-Landau level energy gap. This condition needs to be satisfied for the electronic motion to exhibit topological transport properties of the associated Landau level. In chosen units the inter-Landau level energy gap is unity. (It is equal to $\hbar\w_{c}$, where $\w_{c} = eB/m=1$ is the cyclotron frequency, i.e., the angular frequency of classical cyclotron motion.) Thus, the condition that $V$ is weak and slowly varying is equivalent to the following set of conditions in our chosen units:
%%%%%%%%%%%%%%%%%%%%%%begin%%%%%%%%%%%%%%%%%%%%%%
\begin{align}%\label{eq-}
\pd^{m}_{\mbr}V(\mbr) \ll 1, \quad m = 1, 2\ldots.
\end{align}
%%%%%%%%%%%%%%%%%%%%%%%end%%%%%%%%%%%%%%%%%%%%%%
The condition on $V$ can be used to approximate the operator, $V(\mbr)$, by the first few terms of the Taylor series,
%%%%%%%%%%%%%%%%%%%%%%begin%%%%%%%%%%%%%%%%%%%%%%
\begin{align}\label{eq-expandV}
&V(\mbr) \equiv V(\mbR + \mbpp\times\hat{\mbz})\nn\\
&= V(\mbR) + (\pd_{a}V(\mbR))\le(\e_{ap}\p_{p}\ri) + \frac{1}{2}(\pd^{2}_{ab}V(\mbR))\le(\e_{ap}\e_{bq}\p_{p}\p_{q}\ri)\nn\\
&\quad +  \frac{1}{6}(\pd^{3}_{abc}V(\mbR))\le(\e_{ap}\e_{bq}\e_{cr}\p_{p}\p_{q}\p_{r}\ri) + \ldots.
\end{align}
%%%%%%%%%%%%%%%%%%%%%%%end%%%%%%%%%%%%%%%%%%%%%%
We have used the Einstein convention for summation of repeated indices (over the two values $x$ and $y$) and the two dimensional Levi-Civita tensor $\e_{xx}=\e_{yy} = 0$, $\e_{xy} = - \e_{yx} = 1$. This decomposition has two-fold utility. First, the derivatives of $V$ are all small since $V$ is weak and slowly varying, thus allowing for the use of perturbation theory to calculate modifications to the cyclotron motion. Second, within a Landau level the kinetic momenta, $(\p_{x}, \p_{y})$, are rapidly oscillating relative to each other since they form a canonical pair governed by the SHO Hamiltonian, $K(\mbpp) = (\p_{x}^{2} + \p_{y}^{2})/2$. Consequently, the products of kinetic momentum operators in the above expansion are either rapidly varying or static, thus allowing for a clear separation of time scales.

\subsection{Landau-level projection}\label{sec-LLprojection}

In light of the Taylor expansion in Eq.~\eqref{eq-expandV}, the simplest tractable approximation to the full Hamiltonian, Eq.~\eqref{eq-masterQuadHamil}, involves neglecting all perturbative terms in Eq.~\eqref{eq-expandV}. This crude approximation is equivalent to the standard procedure of Landau-level projection. Within this approximation, the eigenstates of Eq.~\eqref{eq-masterQuadHamil} can be written in the form: 
%%%%%%%%%%%%%%%%%%%%%%begin%%%%%%%%%%%%%%%%%%%%%%
\begin{align}\label{eq-projwfs}
\Psi_{m,n}(\mbR_{x}, \p_{y}) = \c_{m}(R_{x})\y_{n}(\p_{y}), \; E_{m,n} = \vx_{n} + v_{m},
\end{align}
%%%%%%%%%%%%%%%%%%%%%%%end%%%%%%%%%%%%%%%%%%%%%%
where $\y_{n}$ and $\vx_{n}$ are defined in Eq.~\eqref{eq-noFieldSoln}, and the guiding center eigenfunctions are determined by
%%%%%%%%%%%%%%%%%%%%%%begin%%%%%%%%%%%%%%%%%%%%%%
\begin{align}\label{eq-LLprojREigeneqn}
V(\hat{\mbR})\le|\c_{m}\ri\ra = -v_{m}\le|\c_{m}\ri\ra.
\end{align}
%%%%%%%%%%%%%%%%%%%%%%%end%%%%%%%%%%%%%%%%%%%%%%
This eigenequation is unusual because both canonically conjugate operators, $\hat{R}_{x}$ and $\hat{R}_{y}$, are a priori present with equal priority in $V$. However, the nature of the eigenfunctions can be deduced from the well-known operator equations of motion:
%%%%%%%%%%%%%%%%%%%%%%begin%%%%%%%%%%%%%%%%%%%%%%
\begin{align}%\label{eq-}
\dot{\mbR} = i \le[\mc{H}, \mbR\ri] = \hat{\mbz}\times \grad V(\mbr) = - \hat{\mbz}\times \mbE(\mbr),
\end{align}
%%%%%%%%%%%%%%%%%%%%%%%end%%%%%%%%%%%%%%%%%%%%%%
where $\mbE(\mbr) = - \grad V(\mbr)$ is the local electric field. Using our Landau-level projection approximation, $\mbr \sim \mbR$ and so $\dot{\mbR} \simeq - \hat{\mbz}\times \grad V(\mbR)$, which implies that cyclotron orbits drift along equipotentials \cite{2002-yoshioka-zl}. Consequently the (stationary) eigenstates of Eq.~\eqref{eq-LLprojREigeneqn} should also lie along equipotentials and each cover an area of $2\p\ell^{2}$.

%%%%%%%%%%%%%%%%%%%%%%begin%%%%%%%%%%%%%%%%%%%%%%
\begin{figure}[t]
\begin{center}
\resizebox{0.5\textwidth}{!}{\includegraphics{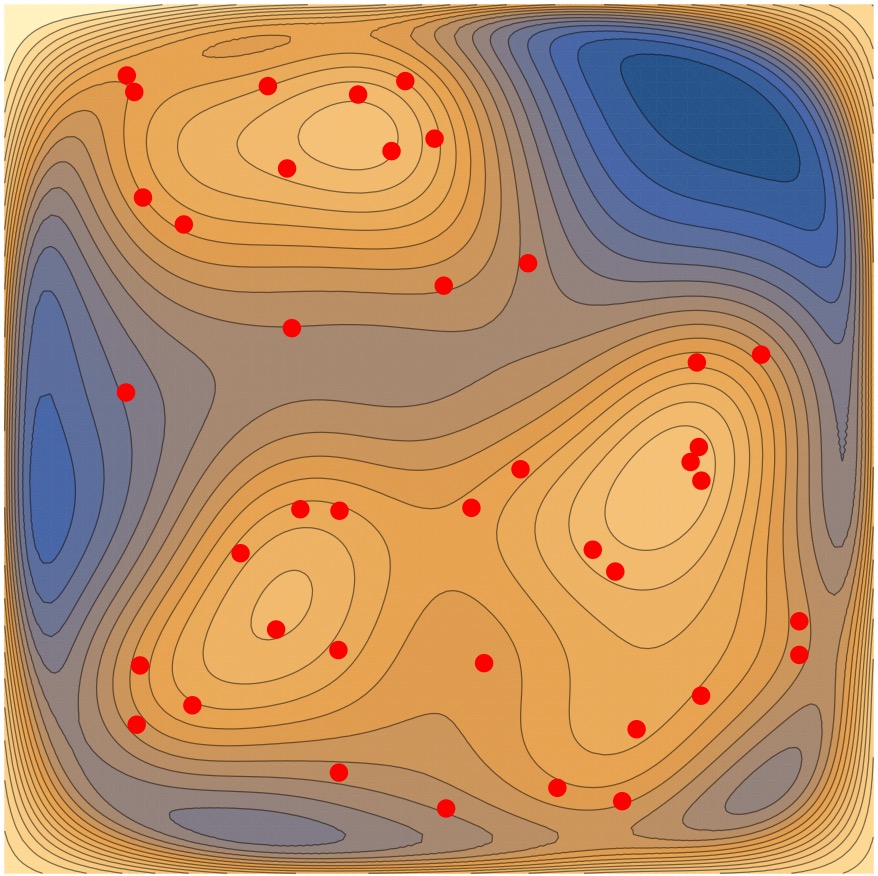}}%inclgphcs[trim=lcm bcm rcm tcm, clip=true, angle=-90]
\caption{Equipotentials in a typical disordered material. The red dots show random potential centers used to simulate this example. These equipotentials are localized and closed in the bulk while those at the boundary run along the entire perimeter of the material. Since guiding center eigenstates run along these equipotentials (see accompanying text), those the bulk are localized. The extended guiding center states along the edge are the well-known edge states, which account for the quantized Hall conductance of a filled Landau level.}
\label{fig-localization}
\end{center}
\end{figure}
%%%%%%%%%%%%%%%%%%%%%%%end%%%%%%%%%%%%%%%%%%%%%%

In the presence of a random disorder potential with short range correlations, equipotentials are closed in the bulk, thus localizing all bulk states as shown in Figure~\ref{fig-localization}. The boundary, assumed to be a steep confining potential, has extended wavefunctions circulating along the perimeter. These are the so-called edge states \cite{1982-halperin-oq}. What we have described here is the standard non-interacting picture of bulk localization in a magnetic field. This accounts for the existence of plateaus in the QHE since the filling up of the bulk localized states does not change macroscopic transport coefficients. The quantized value of Hall conductance can then be derived using standard techniques \cite{2002-yoshioka-zl} when the externally applied chemical potentials at the boundaries differentially populate edge states belonging to the same set of Landau levels.

\subsection{Beyond Landau-level projection - an effective Hamiltonian}\label{sec-beyondLLprojection}

Next we take into account the second and higher order terms in Eq.~\eqref{eq-expandV}. These account for the cyclotron motion induced `jitter' in the electron's position. Since we have assumed that $V(\mbr)$ varies slowly, we can look upon the higher order terms in Eq.~\eqref{eq-expandV} as location-dependent perturbations to the cyclotron orbits. To mathematically articulate this picture we first fix the Landau level index, $n$, and the guiding center location, $\mbR$, and then find the new $\mbR$-dependent kinetic momentum eigenfunction using perturbation theory. Thus, the corrections take the form of a gradient expansion in the electric field; for the purposes of connecting to the gravitational response of quantum Hall states, we need to keep only the first few terms in Eq.~\eqref{eq-expandV}.

Since we are mainly interested in response functions, we only need calculate the first two moments of GIV operators. To this end we will use the following trick to compactly organize the perturbation expansion. Consider a quantum harmonic oscillator Hamiltonian perturbed by operators that are symmetrized polynomials of the canonical pair of momentum and coordinate operators, $(p,q)$. We keep perturbations till the third degree:
%%%%%%%%%%%%%%%%%%%%%%begin%%%%%%%%%%%%%%%%%%%%%%
\begin{align}\label{eq-SHOperturbed}
\mc{H}_{\text{HO}} = \frac{p^{2}+q^{2}}{2} + \e \sum_{k=0}^{3} S_{k},
\end{align}
%%%%%%%%%%%%%%%%%%%%%%%end%%%%%%%%%%%%%%%%%%%%%%
where $S_{k}$ is a real polynomial of degree $k$ in $p$ and $q$, and all terms that mix $p$ and $q$ are symmetrized. For example, the general form of $S_{3}$ is:
%%%%%%%%%%%%%%%%%%%%%%begin%%%%%%%%%%%%%%%%%%%%%%
\begin{align}%\label{eq-}
S_{3} = a p^{3} + b (p^{2}q + p q p + q p^{2}) + c (q^{2}p + q p q + p q^{2}) + d q^{3},\nn
\end{align}
%%%%%%%%%%%%%%%%%%%%%%%end%%%%%%%%%%%%%%%%%%%%%%
where $(a, b, c, d)$ are real numbers. Consider, for a given SHO level index $n$, an effective quadratic Hamiltonian, $\mc{H}_{\text{HO}}^{(n)}$, obtained by Wick-contracting each monomial to terms that are either degree $1$ or $2$ (i.e., linear or quadratic) in $q$ and $p$; each contraction is replaced by the expectation value in the unperturbed energy level with index $n$. For example, following this procedure, $p^{3} \to 3 \le\la p^{2}\ri\ra_{n} p = 3 \vx_{n} p$, where
%%%%%%%%%%%%%%%%%%%%%%begin%%%%%%%%%%%%%%%%%%%%%%
\begin{align}%\label{eq-}
\vx_{n} = n + \frac{1}{2}
\end{align}
%%%%%%%%%%%%%%%%%%%%%%%end%%%%%%%%%%%%%%%%%%%%%%
is the expectation value of $p^{2}$ (or $q^{2}$) in the $(n+1)^{\text{st}}$ energy eigenstate of the unperturbed SHO. Clearly, only $S_{3}$ is modified by this procedure.

The utility of this procedure is that the expectation values of all observables which are linear or quadratic in $q$ and $p$, in the perturbed eigenstate of Eq.~\eqref{eq-SHOperturbed} with index $n$, are obtained correctly to all orders in $S_{0}, S_{1}$ and $S_{2}$, and only to linear order in $S_{3}$, when calculated using the $(n+1)^{\text{st}}$ eigenstate of the effective Hamiltonian, $\mc{H}_{\text{HO}}^{(n)}$. That this is true can be checked by straightforward computation.

This key insight allows us to write down the following effective quadratic Hamiltonian for calculating properties of the $(n+1)^{\text{st}}$ eigenstate of Eq.~\eqref{eq-masterQuadHamil}, using the gradient expansion in Eq.~\eqref{eq-expandV} and keeping up to the third order in derivatives of $V$:
%%%%%%%%%%%%%%%%%%%%%%begin%%%%%%%%%%%%%%%%%%%%%%
\begin{align}%\label{eq-}
\mc{H}^{(n)} &= \frac{\p_{a}\p_{a}}{2} - V(\mbR) - (\pd_{a}V(\mbR))\le(\e_{ap}\p_{p}\ri) \nn\\
&- \frac{1}{2}(\pd^{2}_{ab}V(\mbR))\le(\e_{ap}\e_{bq}\p_{p}\p_{q}\ri) - \nn\\
& \frac{1}{2}(\pd^{3}_{abc}V(\mbR))\le(\e_{ap}\e_{bq}\e_{cr}\vx_{n}\d_{pq}\p_{r}\ri).
\end{align}
%%%%%%%%%%%%%%%%%%%%%%%end%%%%%%%%%%%%%%%%%%%%%%
We have continued to use the Einstein summation convention for repeated indices, and also exploited the commutative property of partial derivatives. Using the identity $\e_{ap}\e_{bq} = \d_{ab}\d_{pq} - \d_{aq}\d_{pb}$, and substituting $\mbE = - \grad V$,
%%%%%%%%%%%%%%%%%%%%%%begin%%%%%%%%%%%%%%%%%%%%%%
\begin{align}%\label{eq-}
\mc{H}^{(n)} &= \frac{\p_{a}\p_{a}}{2} - V(\mbR) \nn\\
&+ \le[E_{a}(\mbR) + \frac{\vx_{n}}{2}\pd_{a}(\grad\cdot\mbE (\mbR))\ri]\le(\e_{ap}\p_{p}\ri) \nn\\
&+ \frac{1}{2}\le[\le(\grad\cdot\mbE (\mbR)\ri)\d_{ab} - \pd_{a}E_{b}(\mbR)\ri]\p_{a}\p_{b}.
\end{align}
%%%%%%%%%%%%%%%%%%%%%%%end%%%%%%%%%%%%%%%%%%%%%%
Note that we have used 2D divergence operators. Consequently, Gauss' law cannot be applied to replace $\grad\cdot\mbE$ with the charge density. Ignoring corrections that are nonlinear in the electric field, the effective Hamiltonian can be compactly expressed as
%%%%%%%%%%%%%%%%%%%%%%begin%%%%%%%%%%%%%%%%%%%%%%
\begin{align}%\label{eq-}
\mc{H}^{(n)} &= \frac{g_{ab}\le(\p_{a} - \Dd_{a}\ri)\le(\p_{b} - \Dd_{b}\ri)}{2} - V(\mbR).
\end{align}
%%%%%%%%%%%%%%%%%%%%%%%end%%%%%%%%%%%%%%%%%%%%%%
In the preceding expression, the new position-dependent `metric' and divergence-free `drift' corrections are, up to third (linear) order in the derivatives of $V$,
%%%%%%%%%%%%%%%%%%%%%%begin%%%%%%%%%%%%%%%%%%%%%%
\begin{subequations}%\label{eq-}
\begin{align}
g_{ab}(\mbR) &= \d_{ab}\le(1+\grad\cdot\mbE (\mbR)\ri) - \pd_{a}E_{b}(\mbR),\label{eq-inducedmetric}\\
\Dd_{a}(\mbR) &= \e_{ab}\le[E_{b}(\mbR) + \frac{\vx_{n}}{2}\pd_{b}(\grad\cdot\mbE (\mbR))\ri].\label{eq-inducedshift}
\end{align}
\end{subequations}
%%%%%%%%%%%%%%%%%%%%%%%end%%%%%%%%%%%%%%%%%%%%%%
As we show below, the determinant of $g$ is responsible for a local modulation in the cyclotron orbit energies, while the unimodular part shears the cyclotron orbit. Thus, we expect the unimodular metric, $G = g/\sqrt{\det g}$, to be the metric relevant for topological response of the quantum Hall state \cite{1992-wen-fk}. A Gaussian curvature field, $K_{G}(\mbR)$, can be extracted from this unimodular metric using the Brioschi formula \cite{2017-abbena-wd}:
%%%%%%%%%%%%%%%%%%%%%%begin%%%%%%%%%%%%%%%%%%%%%%
\begin{align}\label{eq-inducedcurvature}
K_{G}(\mbR) &= - \frac{\grad^{2}\le(\grad_{\mbR}\cdot\mbE(\mbR)\ri)}{4}.
\end{align}
%%%%%%%%%%%%%%%%%%%%%%%end%%%%%%%%%%%%%%%%%%%%%%
We will return to this expression for the curvature in a subsequent section. We note here that an alternate line of reasoning suggests that since $g_{ab}$ appears in the place of an inverse mass matrix, the choice for the spatial metric is the inverse of $G$ used above. This inverse choice will change the sign of the curvature derived above, leaving its magnitude unchanged.

In the presence of a non-uniform electric field the metric is no longer proportional to identity. Instead, the cyclotron orbit is stretched and rotated in a location-dependent manner. To see this note that the metric can be decomposed thus:
%%%%%%%%%%%%%%%%%%%%%%begin%%%%%%%%%%%%%%%%%%%%%%
\begin{align}\label{eq-gdecompose}
g = \sqrt{\det (g)}\le(\L^{-1}\ri)^{T}\L^{-1},
\end{align}
%%%%%%%%%%%%%%%%%%%%%%%end%%%%%%%%%%%%%%%%%%%%%%
where $\L^{-1}$ is a real matrix with unit determinant and composed of a shear and rotations (see below). To the linear third order derivative of $V$,
%%%%%%%%%%%%%%%%%%%%%%begin%%%%%%%%%%%%%%%%%%%%%%
\begin{subequations}%\label{eq-}
\begin{align}
\sqrt{\det (g)}(\mbR) &= 1 + \frac{\grad\cdot\mbE(\mbR)}{2},\\
\L_{ab}(\mbR) &= \le(1- \frac{\grad\cdot\mbE(\mbR)}{4}\ri) \d_{ab} + \frac{\pd_{a} E_{b}}{2}.\label{eq-shearingmatrix}
\end{align}
\end{subequations}
%%%%%%%%%%%%%%%%%%%%%%%end%%%%%%%%%%%%%%%%%%%%%%
Eq.~\eqref{eq-gdecompose} allows us to locally define a rotated and appropriately-rescaled pair of modified kinetic momentum operators:
%%%%%%%%%%%%%%%%%%%%%%begin%%%%%%%%%%%%%%%%%%%%%%
\begin{align}\label{eq-newkineticmom}
\mbs{\Pi} = \L^{-1} \le(\mbpp - \mbs{\Dd}\ri).
\end{align}
%%%%%%%%%%%%%%%%%%%%%%%end%%%%%%%%%%%%%%%%%%%%%%
These operators satisfy the same commutation relations, Eq.~\eqref{eq-GIVcommutation}, as the original kinetic momenta. The guiding center variables, $\mbR$, also need modification to ensure that they commute with the modified kinetic momenta:
%%%%%%%%%%%%%%%%%%%%%%begin%%%%%%%%%%%%%%%%%%%%%%
\begin{align}\label{eq-newguidingcenter}
X_{a} = R_{a} + \le(\pd_{p}\Dd_{a}\ri)\p_{p} - \frac{1}{2} \e_{ar}\e_{ps} \le(\pd_{r}\L_{sq}\ri) \p_{p} \p_{q},
\end{align}
%%%%%%%%%%%%%%%%%%%%%%%end%%%%%%%%%%%%%%%%%%%%%%
where $\Dd$ and $\L$ are evaluated at $\mbR$. To third order in derivatives of $V$, these modified guiding center variables, $\mbX$, commute with the modified kinetic momenta, $\mbs{\Pi}$, and satisfy the GIV commutation relations, Eq.~\eqref{eq-GIVcommutation}.

In terms of these modified GIVs, the effective quadratic Hamiltonian for calculating properties of the $(n+1)^{\text{st}}$ eigenstate of Eq.~\eqref{eq-masterQuadHamil} becomes that of a simple harmonic oscillator with a position-dependent cyclotron frequency:
%%%%%%%%%%%%%%%%%%%%%%begin%%%%%%%%%%%%%%%%%%%%%%
\begin{align}\label{eq-HeffFinal}
\mc{H}^{(n)} &= \w(\mbX)\frac{\mbs{\Pi}^{2}}{2} - V(\mbX),
\end{align}
%%%%%%%%%%%%%%%%%%%%%%%end%%%%%%%%%%%%%%%%%%%%%%
where
%%%%%%%%%%%%%%%%%%%%%%begin%%%%%%%%%%%%%%%%%%%%%%
\begin{align}\label{eq-localcyclotron}
\w(\mbX) = \sqrt{\det (g(\mbX))} = 1 + \frac{\grad\cdot\mbE(\mbX)}{2}.
\end{align}
%%%%%%%%%%%%%%%%%%%%%%%end%%%%%%%%%%%%%%%%%%%%%%
To summarize, the $(n+1)^{\text{st}}$ eigenstate of this Hamiltonian yields the correct linear and quadratic kinetic momentum operator moments, up to linear order in the background potential and the third order in derivatives of $V(\mbR)$ for a fixed value of $\mbR$. The $n$-dependence of this Hamiltonian is hidden in the definitions of the locally-varying parameters and definitions of the altered GIVs. This approach automatically takes into account Landau level mixing by the electric field. For example, the local energy of a particle in the $(n+1)^{\text{st}}$ Landau level is:
%%%%%%%%%%%%%%%%%%%%%%begin%%%%%%%%%%%%%%%%%%%%%%
\begin{align}\label{eq-cyclotronenergy}
E_{n}(\mbR) &= \le\la \mc{H}^{(n)}\ri\ra_{n} = \w(\mbR)\le(n + \frac{1}{2}\ri) - V(\mbR)\nn\\
&= \le(1 + \frac{\grad\cdot\mbE(\mbR)}{2}\ri)\le(n + \frac{1}{2}\ri) - V(\mbR).
\end{align}
%%%%%%%%%%%%%%%%%%%%%%%end%%%%%%%%%%%%%%%%%%%%%%
The local Landau level spacing is thus modified by a non-uniform electric field. This observable effect takes into account Landau-level-mixing by the electric field and was predicted earlier in \cite{1987-fertig-rt,1997-haldane-sf}.

Before proceeding to calculate the response of other observables to the non-uniform electric field, we use the effective Hamiltonian in Eq.~\eqref{eq-HeffFinal} to derive a simple geometric picture for the effect of the electric field on the cyclotron orbits. (See Figure~\ref{fig-cyclotronorbit}.) Clearly, the form of Eq.~\eqref{eq-HeffFinal} implies that the modified cyclotron orbits are circular in $\mbs{\Pi}$-space. The $\Pi$ coordinates were obtained from the original kinetic momenta via the linear transformation, Eq.~\eqref{eq-newkineticmom}, which is the combination of a shift by $\mbs{\Dd}$ and a unimodular transformation, $\L^{-1}$. The transformation $\L$ can be decomposed \cite{2011-gosson-oq} as: 
%%%%%%%%%%%%%%%%%%%%%%begin%%%%%%%%%%%%%%%%%%%%%%
\begin{align}\label{eq-orbitellipticity}
\L = R(-\th)\cdot \le(\ba{cc} \l & 0 \\ 0 & \l^{-1} \ea\ri)\cdot R(\th),
\end{align}
%%%%%%%%%%%%%%%%%%%%%%%end%%%%%%%%%%%%%%%%%%%%%%
where $\th$ is the angle by which the coordinate axes need to be rotated to ensure that $\pd_{x}E_{x}-\pd_{y}E_{y}$ is maximized. $\l = 1 + \le(\pd_{x}E_{x}-\pd_{y}E_{y}\ri)_{\text{max}}/2$, where the derivatives are evaluated in the new orientation specified by $\th$. Thus, the cyclotron orbits in real space are sheared, with the long axis aligned with the $x$-axis of the rotated coordinate frame in which $\pd_{x}E_{x}-\pd_{y}E_{y}$ is maximum. The ratio of the two axes of the elliptical orbit is given by $\l^{2}$. There are two additional modifications to the field-free cyclotron orbits. First, an orbit at the original field-free location $\mbR$ is translated by an amount $\mbs{\d}(\mbR) = \le\la \mbr - \mbR \ri\ra = \hat{\mbz} \times \mbs{\Dd}(\mbR)$. Second, these orbits are no longer stationary and acquire a drift velocity $\mbv_{d} = \le\la \mbpp \ri\ra = \mbs{\Dd}(\mbR)$ which is perpendicular to the shift, $\mbs{\d}(\mbR)$. Figure~\ref{fig-cyclotronorbit} summarizes these changes in the geometry of cyclotron orbits, when placed in an external electric field. 

\subsection{Local observables in a non-uniform electric field}\label{sec-observablecalculation}

Now we consider how different observable quantities change when a non-uniform field is switched on. Using our geometric picture we can delineate these changes as arising due to (i) displacement and drift i.e., (due to $\mbs{\Dd}$) and (ii) shearing of the orbits (due to an effective distorted real space metric, whose effect is encapsulated in the matrix $\L$).

These calculations are succinct using the Wigner pseudoprobability formalism \cite{1949-moyal-vn}. The central idea is to replace the quantum wavefunction, which is a function of one coordinate from each independent canonically conjugate pair of variables, by the Wigner pseudoprobability distribution, which is defined over the entire canonical phase space.
%%%%%%%%%%%%%%%%%%%%%%begin%%%%%%%%%%%%%%%%%%%%%%
\begin{align}%\label{eq-}
W_{\Psi}(\mbR,\mbp) &= \iint \frac{dR}{2\p}\frac{d\p}{2\p} \Psi^{*}(R_{x}+R/2,\p_{y}+\p/2)\times\nn\\
&\qquad\Psi(R_{x}-R/2,\p_{y}-\p/2)e^{i(R_{y}R+\p_{x}\p)}.
\end{align}
%%%%%%%%%%%%%%%%%%%%%%%end%%%%%%%%%%%%%%%%%%%%%%

This formalism provides a natural framework for calculations involving GIVs, since typical observables expressed using GIVs do not favor any particular component in the canonical pair of guiding center coordinates. Given an operator $\hat{\mc{O}}(\mbR,\mbpp)$, where the products of canonically conjugate variables have been symmetrized, the expectation value of $\hat{\mc{O}}$ in state $\Psi$ is found by simply integrating the product of the Wigner function and the \emph{classical} function $\mc{O}(\mbR,\mbpp)$ over the $R_{x}-R_{y}-\p_{x}-\p_{y}$ phase space.

Within the scheme of Landau-level projection, the Wigner function corresponding to the product wavefunctions in \eqref{eq-projwfs} is also a product of Wigner functions in the guiding center and kinetic momenta phase spaces:
%%%%%%%%%%%%%%%%%%%%%%begin%%%%%%%%%%%%%%%%%%%%%%
\begin{align}%\label{eq-}
\Psi_{m,n}(R_{x}, \p_{y}) &= \c_{m}(R_{x})\y_{n}(\p_{y}) \nn\\
\Leftrightarrow W_{m,n}(\mbR,\mbp) &= \mc{W}_{m}(\mbR)w_{n}(\mbpp).\nn
\end{align}
%%%%%%%%%%%%%%%%%%%%%%%end%%%%%%%%%%%%%%%%%%%%%%
Using the effective Hamiltonian, Eq.~\eqref{eq-HeffFinal}, we conclude that the energy eigenstates are still of the form \eqref{eq-projwfs}, except that they are functions of the modified $\mbs{\Pi}-\mbX$ phase space coordinate pairs (which were defined in Eqs.~\eqref{eq-newkineticmom} and \eqref{eq-newguidingcenter}):
%%%%%%%%%%%%%%%%%%%%%%begin%%%%%%%%%%%%%%%%%%%%%%
\begin{align}%\label{eq-}
W_{m,n}(\mbR,\mbp) &= \mc{W}_{m}(\mbX)w_{n}(\mbs{\Pi}).
\end{align}
%%%%%%%%%%%%%%%%%%%%%%%end%%%%%%%%%%%%%%%%%%%%%%
Since the bulk guiding center wavefunctions, $\c_{m}$, form a complete basis, we also have the following completeness relation, correct to the third linear order in derivatives of $V$:
%%%%%%%%%%%%%%%%%%%%%%begin%%%%%%%%%%%%%%%%%%%%%%
\begin{align}\label{eq-guidingcompleteness}
\sum_{m}\mc{W}_{m}(\mbR) = \frac{1}{2\p} = \sum_{m}\mc{W}_{m}(\mbX).
\end{align}
%%%%%%%%%%%%%%%%%%%%%%%end%%%%%%%%%%%%%%%%%%%%%%

\subsubsection{Local current density - analytical approach}

Following Eq.~\eqref{eq-localopsinGIV}, we consider the local single particle charge current density at location $\mbx$,
%%%%%%%%%%%%%%%%%%%%%%begin%%%%%%%%%%%%%%%%%%%%%%
\begin{align}%\label{eq-}
\hat{\mbj}(\mbx) = -\frac{1}{2}\le\{\hat{\mbpp}, \d\le(\hat{\mbr} - \mbx\ri)\ri\}.
\end{align}
%%%%%%%%%%%%%%%%%%%%%%%end%%%%%%%%%%%%%%%%%%%%%%
Carets denote operators and we have used the fact that for quadratic dispersion the velocity operator is simply the kinetic momentum.

Within a single filled Landau level with index $n$, the sum of the expectation values of this operator in all single-particle states yields the total local current density, $\mbj^{(n)}(\mbx)$. In component notation
%%%%%%%%%%%%%%%%%%%%%%begin%%%%%%%%%%%%%%%%%%%%%%
\begin{align}%\label{eq-}
j_{a}^{(n)}(\mbx) &= -\sum_{m} \iint d^{2}Xd^{2}\Pi \; W_{m,n}(\mbX,\mbs{\Pi}) \p_{a}\d^{2}(\mbr-\mbx)\nn\\
&= -\iint \frac{d^{2}Xd^{2}\Pi}{2\p} \; w_{n}(\mbs{\Pi}) \p_{a}\d^{2}(\mbr-\mbx).
\end{align}
%%%%%%%%%%%%%%%%%%%%%%%end%%%%%%%%%%%%%%%%%%%%%%
The completeness relation, Eq.~\eqref{eq-guidingcompleteness}, was used to eliminate the guiding center Wigner function. In this expression, $\mbpp$ and $\mbr$ are functions of $\mbX$ and $\mbs{\Pi}$, as defined in Eqs.~\eqref{eq-newkineticmom} and \eqref{eq-newguidingcenter}. Substituting these expressions,
%%%%%%%%%%%%%%%%%%%%%%begin%%%%%%%%%%%%%%%%%%%%%%
\begin{align}\label{eq-deltafuncref}
j_{a}^{(n)}(\mbx) &= -\iint \frac{d^{2}Xd^{2}\Pi}{2\p} \; w_{n}(\mbs{\Pi}) (\L \mbs{\Pi} + \mbs{\Dd})_{a} \nn\\
& \quad\quad\quad \times \d^{2}\le[\mbr(\mbs{\Pi},\mbX)-\mbx\ri],
\end{align}
%%%%%%%%%%%%%%%%%%%%%%%end%%%%%%%%%%%%%%%%%%%%%%
where
%%%%%%%%%%%%%%%%%%%%%%begin%%%%%%%%%%%%%%%%%%%%%%
\begin{align}%\label{eq-}
r_{a}(\mbs{\Pi},\mbX) &= X_{a} + \e_{ab} \Dd_{b}(\mbX) + \le[\e_{ac}\L_{cb}(\mbX) - \pd_{b}\Dd_{a}(\mbX)\ri]\Pi_{b}\nn\\
& \qquad + \frac{1}{2}  \e_{ar}\e_{ps} \le(\pd_{r}\L_{sq}(\mbX)\ri) \Pi_{p} \Pi_{q}.
\end{align}
%%%%%%%%%%%%%%%%%%%%%%%end%%%%%%%%%%%%%%%%%%%%%%
The delta function in Eq.~\eqref{eq-deltafuncref} has a zero at $\mbX = \tilde{\mbX}$:
%%%%%%%%%%%%%%%%%%%%%%begin%%%%%%%%%%%%%%%%%%%%%%
\begin{align}\label{eq-orbitfromlocation}
\tilde{X}_{a}(\mbx) &= x_{a} - \e_{ab}\Pi_{b} + \ldots,
\end{align}
%%%%%%%%%%%%%%%%%%%%%%%end%%%%%%%%%%%%%%%%%%%%%%
where the ellipsis denote terms which are of the same order of smallness as the electric field. Thus, for any function $F$,
%%%%%%%%%%%%%%%%%%%%%%begin%%%%%%%%%%%%%%%%%%%%%%
\begin{align}%\label{eq-}
 \iint d^{2}X \d^{2}(\mbr-\mbx) F(\mbX) = J^{-1}(\mbx) F(\tilde{\mbX}(\mbx)),\nn
\end{align}
%%%%%%%%%%%%%%%%%%%%%%%end%%%%%%%%%%%%%%%%%%%%%%
where $J(\mbx)$ is the Jacobian arising from the delta function integral:
%%%%%%%%%%%%%%%%%%%%%%begin%%%%%%%%%%%%%%%%%%%%%%
\begin{align}%\label{eq-}
&J(\mbx) = \le|\det \le(\frac{\pd r_{a}}{\pd X_{b}}\ri)\ri|_{\mbX = \tilde{\mbX}(\mbx)}\nn\\
&= \le[\e_{ac}\pd_{a}\Dd_{c}(\mbX) + \e_{ac}\Pi_{p}\pd_{a}\L_{cp}(\mbX)\ri]_{\mbX = \tilde{\mbX}(\mbx)}\nn\\
&= 1 + \e_{ac}\pd_{a}\Dd_{c}(\mbx) - \le(\pd^{2}_{aa}\Dd_{p}(\mbx) - \e_{ac}\pd_{a}\L_{cp}(\mbx)\ri)\Pi_{p} +\nn\\
& \qquad \le(\frac{\e_{ab}\e_{cp}\e_{dq}}{2}\pd^{3}_{acd}\Dd_{b} - \e_{ac}\e_{bq}\pd^{2}_{ab}\L_{cp}\ri)\Pi_{p}\Pi_{q} .
\end{align}
%%%%%%%%%%%%%%%%%%%%%%%end%%%%%%%%%%%%%%%%%%%%%%
The above expression is correct to third order in derivatives of $\mbE$. Next,
%%%%%%%%%%%%%%%%%%%%%%begin%%%%%%%%%%%%%%%%%%%%%%
\begin{align}%\label{eq-}
&j_{a}^{(n)}(\mbx) \nn\\
&= -\iint \frac{d^{2}\Pi}{2\p}\; w_{n}(\mbs{\Pi}) (\L(\tilde{\mbX}(\mbx)) \mbs{\Pi} + \mbs{\Dd}(\tilde{\mbX}(\mbx)))_{a} J^{-1}(\mbx)\nn\\
&= -\frac{\Dd_{a}(\mbx)}{2\p} - \le(- \e_{de}\pd_{d}\L_{ac} + \frac{1}{2}\e_{bc}\e_{de}\pd^{2}_{bd}\Dd_{a}\ri)\frac{\le\la \Pi_{c} \Pi_{e}\ri\ra_{n}}{2\p}\nn\\
& \qquad + \le(\e_{dc} \pd_{d}\L_{cb} - \pd^{2}_{dd} \Dd_{b}\ri)\frac{\le\la \Pi_{a} \Pi_{b}\ri\ra_{n}}{2\p}\nn\\
&= -\frac{\Dd_{a}(\mbx)}{2\p} - \frac{\vx_{n}}{2\p} \le(\frac{3}{2}\pd^{2}_{dd}\Dd_{a} - 2 \e_{dc}\pd_{d}\L_{ac}\ri).
\end{align}
%%%%%%%%%%%%%%%%%%%%%%%end%%%%%%%%%%%%%%%%%%%%%%
The current density can be separated into two contributions. First, a contribution arising from the drift-displacement vector, $\mbs{\Dd}$:
%%%%%%%%%%%%%%%%%%%%%%begin%%%%%%%%%%%%%%%%%%%%%%
\begin{align}%\label{eq-}
\le[\mbj_{\Dd}^{(n)}(\mbx)\ri]_{a} &= -\frac{\Dd_{a}(\mbx)}{2\p} - \frac{3}{2}\pd^{2}_{dd}\Dd_{a}\frac{\vx_{n}}{2\p}\nn\\
&=  -\frac{\e_{ab}}{2\p}\le[\mbE + 2 \vx_{n}\grad\le(\grad\cdot\mbE\ri)\ri]_{b}.
\end{align}
%%%%%%%%%%%%%%%%%%%%%%%end%%%%%%%%%%%%%%%%%%%%%%
Second, another contribution involving the shear matrix, $\L$:
%%%%%%%%%%%%%%%%%%%%%%begin%%%%%%%%%%%%%%%%%%%%%%
\begin{align}\label{eq-}
\le[\mbj_{\L}^{(n)}(\mbx)\ri]_{a} = 2 \e_{dc}\pd_{d}\L_{ac}(\mbx) \frac{\vx_{n}}{2\p} = \frac{\e_{ab}}{2\p}\frac{\vx_{n}}{2}\le[\grad\le(\grad\cdot\mbE\ri)\ri]_{b}.
\end{align}
%%%%%%%%%%%%%%%%%%%%%%%end%%%%%%%%%%%%%%%%%%%%%%
Adding these, we obtain the total current density contributed by a filled Landau level with index $n$:
%%%%%%%%%%%%%%%%%%%%%%begin%%%%%%%%%%%%%%%%%%%%%%
\begin{align}\label{eq-totalcurrentdensity}
j_{a}^{(n)}(\mbx) &= \le[\mbj_{\Dd}^{(n)}(\mbx)\ri]_{a} + \le[\mbj_{\L}^{(n)}(\mbx)\ri]_{a} \nn\\
&= -\frac{\e_{ab}}{2\p}\le[\mbE + \frac{3\vx_{n}}{2}\grad\le(\grad\cdot\mbE\ri)\ri]_{b}.
\end{align}
%%%%%%%%%%%%%%%%%%%%%%%end%%%%%%%%%%%%%%%%%%%%%%
The preceding expression for linear response is correct up to the third order in derivatives of the electrostatic potential and agrees with previous derivations \cite{1993-simon-qq,2012-hoyos-fk}.

It is known from field theoretical approaches \cite{2012-hoyos-fk,2009-read-fk} that the coefficient of the second term in Eq.~\eqref{eq-totalcurrentdensity} arises from a combination of Hall viscosity and a term that originates from the swirling motion of cyclotron orbits \cite{2012-hoyos-fk}. Since the motion of cyclotron orbits is given by the drift velocity, $\mbv_{d} = \mbs{\Dd}$, which involves only $\mbs{\Dd}$. We formulate the following conjecture: the Hall viscosity contribution (which is related to the gravitational coupling constant of quantum Hall states \cite{1992-wen-fk,2009-read-fk,2012-hoyos-fk}) is given by $\mbj_{\L}^{(n)}(\mbx)$, the current density arising from the shearing of cyclotron orbits. The magnitude of $\mbj_{\L}^{(n)}(\mbx)$ matches that obtained from the Hall viscosity contribution, thus yielding the correct values for the Hall viscosity and gravitational response coefficients.

\subsubsection{Local charge density - analytical approach}

Following Eq.~\eqref{eq-localopsinGIV}, the single particle charge density operator at location $\mbx$ is
%%%%%%%%%%%%%%%%%%%%%%begin%%%%%%%%%%%%%%%%%%%%%%
\begin{align}%\label{eq-}
\hat{\rr}(\mbx) = - \d\le(\hat{\mbr} - \mbx\ri).
\end{align}
%%%%%%%%%%%%%%%%%%%%%%%end%%%%%%%%%%%%%%%%%%%%%%
Within a single filled Landau level with index $n$, the sum of the expectation values of this operator in all single-particle states yields the total local charge density, $\rr^{(n)}(\mbx)$. Using techniques introduced previously for calculating the current density operator, we find:
%%%%%%%%%%%%%%%%%%%%%%begin%%%%%%%%%%%%%%%%%%%%%%
\begin{align}%\label{eq-}
\rr^{(n)}(\mbx) &= -\sum_{m} \iint d^{2}Xd^{2}\Pi \; W_{m,n}(\mbX,\mbs{\Pi}) \d^{2}(\mbr-\mbx)\nn\\
&= -\iint \frac{d^{2}Xd^{2}\Pi}{2\p} \; w_{n}(\mbs{\Pi}) \d^{2}(\mbr-\mbx)\nn\\
&= -\iint \frac{d^{2}\Pi}{2\p}\; w_{n}(\mbs{\Pi}) J^{-1}(\mbx)\nn\\
&= - \frac{1}{2\p}\bigg[1 - \e_{ac}\pd_{a}\Dd_{c}(\mbx) - \nn\\
&\quad \vx_{n} \le(\frac{\grad^{2}\le(\e_{ab}\pd_{a}\Dd_{b}\ri)}{2} - \e_{ac}\e_{bp}\pd^{2}_{ab}\L_{cp}\ri)\bigg].
\end{align}
%%%%%%%%%%%%%%%%%%%%%%%end%%%%%%%%%%%%%%%%%%%%%%
The expression above is correct to third order in derivatives of the electric field. Using $\rr_{0} = - (2\p)^{-1}$ to denote charge density in the absence of electric field, the contributions from $\mbs{\Dd}$ and $\L$ can be separated as follows:
%%%%%%%%%%%%%%%%%%%%%%begin%%%%%%%%%%%%%%%%%%%%%%
\begin{subequations}\label{eq-chargeresponsepartial}
\begin{align}
\le[\frac{\rr^{(n)}(\mbx) - \rr_{0}}{\rr_{0}}\ri]_{\Dd} &= - \e_{ac}\pd_{a}\Dd_{c}(\mbx) - \frac{\vx_{n}}{2}\grad^{2}\le(\e_{ab}\pd_{a}\Dd_{b}\ri) \nn\\
&= \grad\cdot\mbE (\mbx) + \vx_{n} \grad^{2}\le(\grad\cdot\mbE\ri) + \ldots\label{eq-chargeresponsepartialdelta}\\
\le[\frac{\rr^{(n)}(\mbx) - \rr_{0}}{\rr_{0}}\ri]_{\L} &= \vx_{n}\e_{ac}\e_{bp}\pd^{2}_{ab}\L_{cp} \nn\\
&= - \frac{\vx_{n}}{4} \grad^{2}\le(\grad\cdot\mbE\ri) + \ldots \label{eq-chargeresponsepartialshear}
\end{align}
\end{subequations}
%%%%%%%%%%%%%%%%%%%%%%%end%%%%%%%%%%%%%%%%%%%%%%
Adding these contributions, the local fractional change in charge density becomes:
%%%%%%%%%%%%%%%%%%%%%%begin%%%%%%%%%%%%%%%%%%%%%%
%\begin{subequations}
\begin{align}\label{eq-chargeresponse}
\frac{\rr^{(n)}(\mbx) - \rr_{0}}{\rr_{0}} &= \grad\cdot\mbE (\mbx) + \frac{3\vx_{n}}{4} \grad^{2}\le(\grad\cdot\mbE\ri) + \ldots.
\end{align}
%\end{subequations}
%%%%%%%%%%%%%%%%%%%%%%%end%%%%%%%%%%%%%%%%%%%%%%
Our result is consistent with previous calculations on the linear response of quantum Hall states \cite{1993-simon-qq,2014-abanov-rm}.

The contribution (Eq.~\eqref{eq-chargeresponsepartialdelta}) from the orbital shift field, $\mbs{\Dd}$, can be interpreted simply in terms of the geometric picture sketched in Figure~\ref{fig-cyclotronorbit}. The non-uniform electric field causes the orbit at location $\mbX$ to shift by an amount $\mbs{\d}(\mbX) = \hat{\mbz} \times \mbs{\Dd}(\mbX)$. Ignoring the effects of orbit shear, this induces a coarse-grained charge polarization field $\mbP(\mbX) = \rr_{0} \mbs{\d}(\mbX)$. This polarization field induces an excess charge $\rr(\mbx) - \rr_{0} = - \le\la \grad\cdot \mbP(\tilde{\mbX}(\mbx))\ri\ra$, a standard result in the study of electrostatics in continuous media; the angular brackets denote an average over orbits which contribute to the charge at $\mbx$ while $\tilde{\mbX}(\mbx)$ was defined in Eq.~\eqref{eq-orbitfromlocation}. Expressing $\mbP$ in terms of $\mbs{\Dd}$ and using Eq.~\eqref{eq-orbitfromlocation}, we arrive at the $\mbs{\Dd}$-contributions in Eqs.~\eqref{eq-chargeresponsepartialdelta} and \eqref{eq-chargeresponse}.

We can obtain the contribution (Eq.~\eqref{eq-chargeresponsepartialshear}) from the shearing field, $\L$, by exploiting an apparently unrelated property of quantum Hall states. It is known that spatial curvature induces excess charge in quantum Hall states \cite{1992-wen-fk}, a phenomenon we term topological gravitational response. For the fully filled Landau level with index $n$, this becomes:
%%%%%%%%%%%%%%%%%%%%%%begin%%%%%%%%%%%%%%%%%%%%%%
\begin{align}\label{eq-tgr}
\d \rr_{G}(\mbx) = -\frac{\k}{2\p} K_{G}(\mbx) \equiv \rr_{0} \vx_{n} K_{G}(\mbx),
\end{align}
%%%%%%%%%%%%%%%%%%%%%%%end%%%%%%%%%%%%%%%%%%%%%%
where $K_{G}(\mbx)$ is the local Gaussian curvature and $\d \rr_{G}$ denotes the change in charge density arising due to topological gravitational response. $\vx_{n}$ is the value of the gravitational coupling constant, $\k$, associated with a fully-filled Landau level with index $n$. $\k$ is believed to be a topologically-protected quantity which can only take up rational fraction values.

While there is no literal real-space curvature in the scenario we are considering, we have already noted that the non-uniform electric field can induce a fictitious Gaussian curvature, given by Eq.~\eqref{eq-inducedcurvature}. Therefor, we conjecture that the introduction of this curvature has the same effect as that of a spatial curvature with the same magnitude. Then it follows that the physics of topological gravitational response contributes the following amount to the induced charge:
%%%%%%%%%%%%%%%%%%%%%%begin%%%%%%%%%%%%%%%%%%%%%%
\begin{align}%\label{eq-}
\d \rr_{G}(\mbx) = - \rr_{0} \frac{\vx_{n}}{4} \grad^{2}\le(\grad\cdot\mbE\ri).
\end{align}
%%%%%%%%%%%%%%%%%%%%%%%end%%%%%%%%%%%%%%%%%%%%%%
This is exactly the value obtained from the shear contribution, Eq.~\eqref{eq-chargeresponsepartialshear}. We have thus shown that topological gravitational response apparently contributes to the local charge density response of quantum Hall states in non-uniform electric fields.

We conclude this section with the following conjecture, in analogy with the connection between the current density response and the gravitational coupling constant \cite{2012-hoyos-fk,2009-read-fk}. We expect that the charge density response to a non-uniform electric field should have the form:
%%%%%%%%%%%%%%%%%%%%%%begin%%%%%%%%%%%%%%%%%%%%%%
\begin{align}\label{eq-chargeresponseconjecture}
\rr(\mbx) - \rr_{0} = - \grad\cdot \mbP(\mbx) + \frac{\k}{2\p} \frac{\grad^{2}\le(\grad\cdot\mbE\ri)}{4} + \ldots,
\end{align}
%%%%%%%%%%%%%%%%%%%%%%%end%%%%%%%%%%%%%%%%%%%%%%
where $\rr_{0}$ is the charge density in the absence of any electric field, $\k$ is the gravitational coupling constant and $\mbP(\mbx)$ is the averaged polarization field caused by shifts in the guiding centers.

%%%%%%%%%%%%%%%%%%%%%%begin%%%%%%%%%%%%%%%%%%%%%%
\begin{figure}[t]
\begin{center}
\resizebox{0.5\textwidth}{!}{\includegraphics{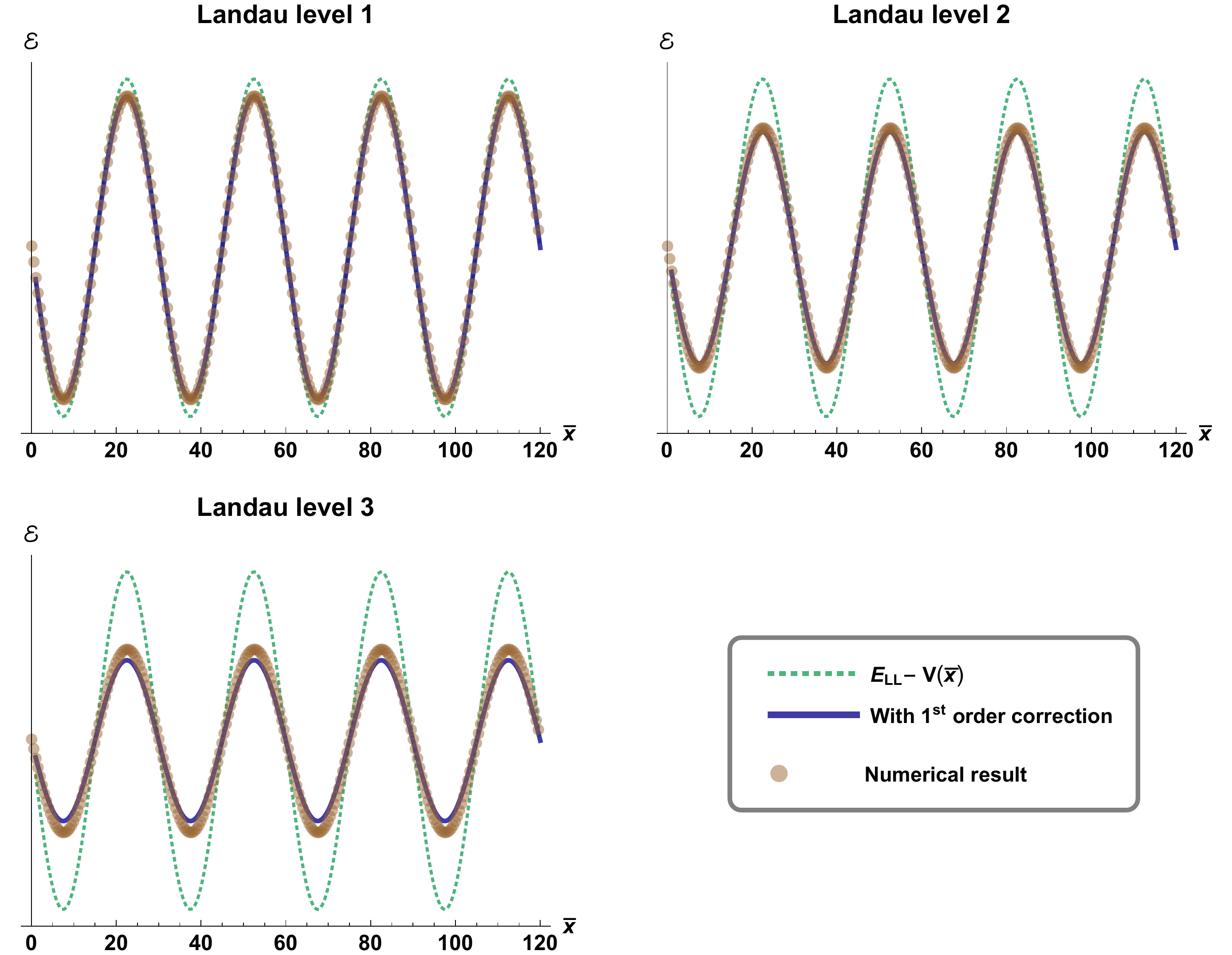}}%inclgphcs[trim=lcm bcm rcm tcm, clip=true, angle=-90]
\caption{The variation of cyclotron orbit energy with orbit location, $\ol{x}$. The Landau levels are modeled by the lowest bands in a Hofstadter model on a square lattice. The nonuniform electric field is generated by a sinusoidal background potential, which is small compared to the inter-Landau-level energy gap, so that the system in the linear response regime. The brown circles correspond to the cyclotron energies obtained via numerical diagonalization. The dashed green curve is the sum of the Landau level energy and the local potential energy, which is the correct energy when the electric field is uniform. The thick blue curve corresponds to Eqs.~\eqref{eq-cyclotronenergy} and \eqref{eq-cyclotronenergy1d}, correct up to the second order in the derivatives of the electric field. For these plots, $V_{0}/\e_{c} = 0.05$, $k\ell = 0.65$.}
\label{fig-enplotgrid}
\end{center}
\end{figure}
%%%%%%%%%%%%%%%%%%%%%%%end%%%%%%%%%%%%%%%%%%%%%%

\subsubsection{Numerical checks of analytical calculations}

Now we provide numerical checks for our analytical results on how local observables change as a function of a spatially-varying electric field. To this end we construct a Hofstadter model on a square lattice with nearest-neighbor hopping and periodic boundary conditions in the $x$-direction. To this we add a sinusoidally-varying on-site electrostatic potential, $V(x) = V_{0} \sin k x$. We use the Landau gauge, $\mbA = B x \hat{\mby}$, yielding eigenstates which are extended in the $y$-direction but localized in the $x$-direction. In this system, it is natural to set the hopping amplitude, the lattice spacing, Planck's constant, $h$, and the magnitude of electronic charge, $e$, to unity. In these units, we choose the magnetic field to be $B=1/q$, where $q\gg1$. With this choice, the lowest few Landau levels have the same characteristics as obtained for a continuum model with quadratically-dispersing particles. Diagonalizing the Hamiltonian, we obtain the spatial variation of cyclotron orbit energies, the local charge-current density and the charge density.

%%%%%%%%%%%%%%%%%%%%%%begin%%%%%%%%%%%%%%%%%%%%%%
\begin{figure}[t]
\begin{center}
\resizebox{0.5\textwidth}{!}{\includegraphics{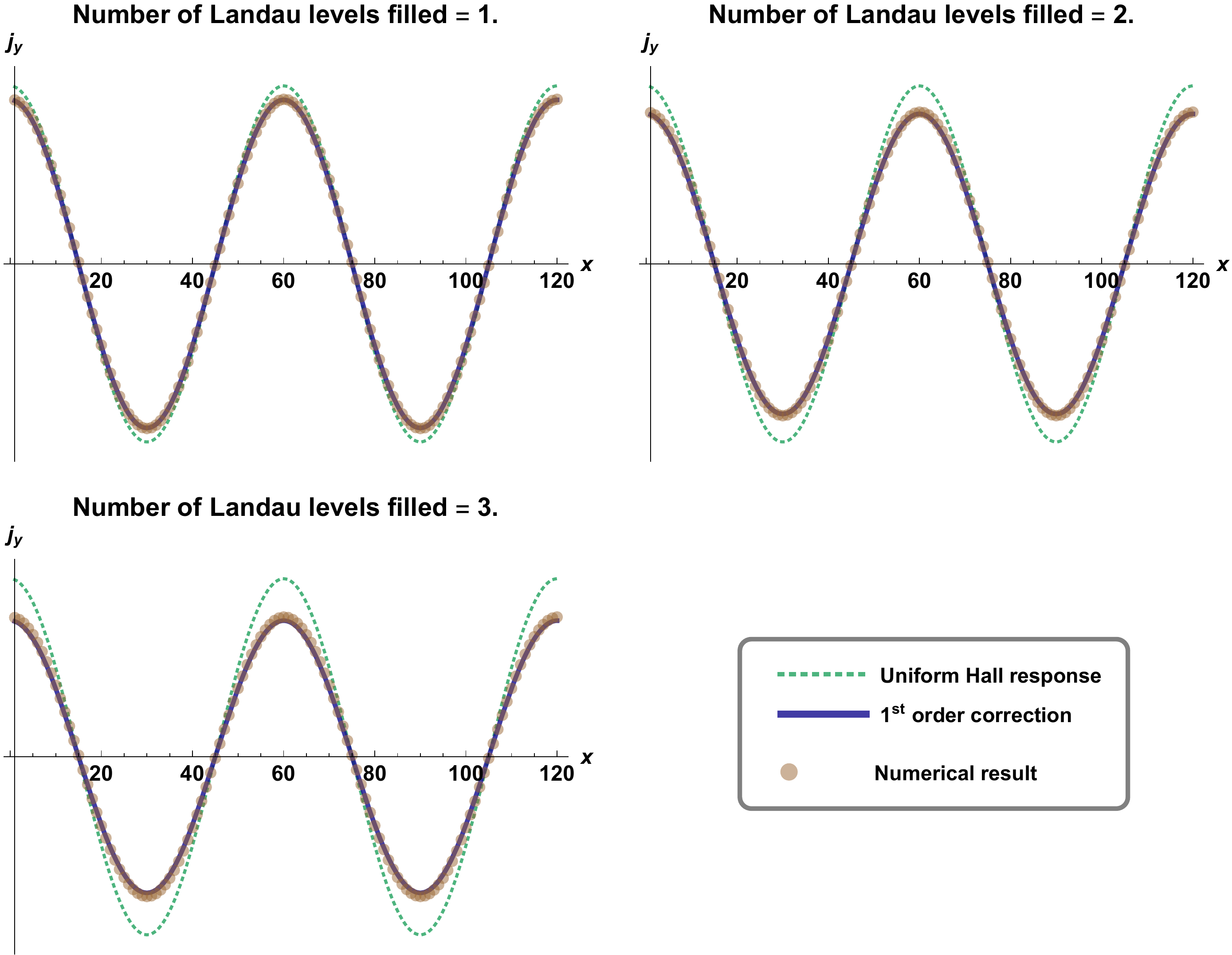}}%inclgphcs[trim=lcm bcm rcm tcm, clip=true, angle=-90]
\caption{The spatial variation of local current density in a nonuniform (sinusoidal) electric field. The Landau levels are modeled by the lowest bands in a Hofstadter model on a square lattice. The nonuniform electric field is generated by a sinusoidal background potential, which is small compared to the inter-Landau-level energy gap so that the system in the linear response regime. The brown circles correspond to the local current density values obtained via numerical diagonalization. The dashed green curve is the quantized local Hall response, which is correct when the electric field is uniform. The thick blue curve corresponds to Eqs.~\eqref{eq-totalcurrentdensity} and \eqref{eq-totalcurrentdensity1d}, correct up to the second order in the derivatives of the electric field. For these plots, $V_{0}/\e_{c} = 0.05$, $k\ell = 0.32$.}
\label{fig-curplotgrid}
\end{center}
\end{figure}
%%%%%%%%%%%%%%%%%%%%%%%end%%%%%%%%%%%%%%%%%%%%%%

Below, we use slightly modified units better-suited for the Hofstadter model: the units of length and energy are respectively set to the lattice spacing and the nearest-neighbor hopping amplitude. In these units the magnetic length is $\ell = \sqrt{q/2\p}$ and the inter-Landau level (cyclotron) gap, in the continuum limit, is $\e_{c} = 4\p/q$. For $k\ell \gg 1$ (slowly varying potential) and $V_{0} \ll \e_{c}$ (weak potential) our results can be written as follows.

The cyclotron orbit energies (from Eq.~\eqref{eq-cyclotronenergy}) become the local energy of each quantum state:
%%%%%%%%%%%%%%%%%%%%%%begin%%%%%%%%%%%%%%%%%%%%%%
\begin{align}\label{eq-cyclotronenergy1d}
E_{n}(\ol{x}) &=  \le(n + \frac{1}{2}\ri) - \le(1 - \frac{(k\ell)^{2}}{2}\le(n + \frac{1}{2}\ri) + \ldots \ri) V_{0} \sin k \ol{x}.
\end{align}
%%%%%%%%%%%%%%%%%%%%%%%end%%%%%%%%%%%%%%%%%%%%%%
In this expression, $\ol{x}$ denotes the average $x$-position of the quantum state and $n$ denotes the index of the Landau level to which the orbit belongs. In Figure~\ref{fig-enplotgrid}, we have shown the numerical verification for this relation for the lowest three Landau levels.

%%%%%%%%%%%%%%%%%%%%%%begin%%%%%%%%%%%%%%%%%%%%%%
\begin{figure}[t]
\begin{center}
\resizebox{0.5\textwidth}{!}{\includegraphics{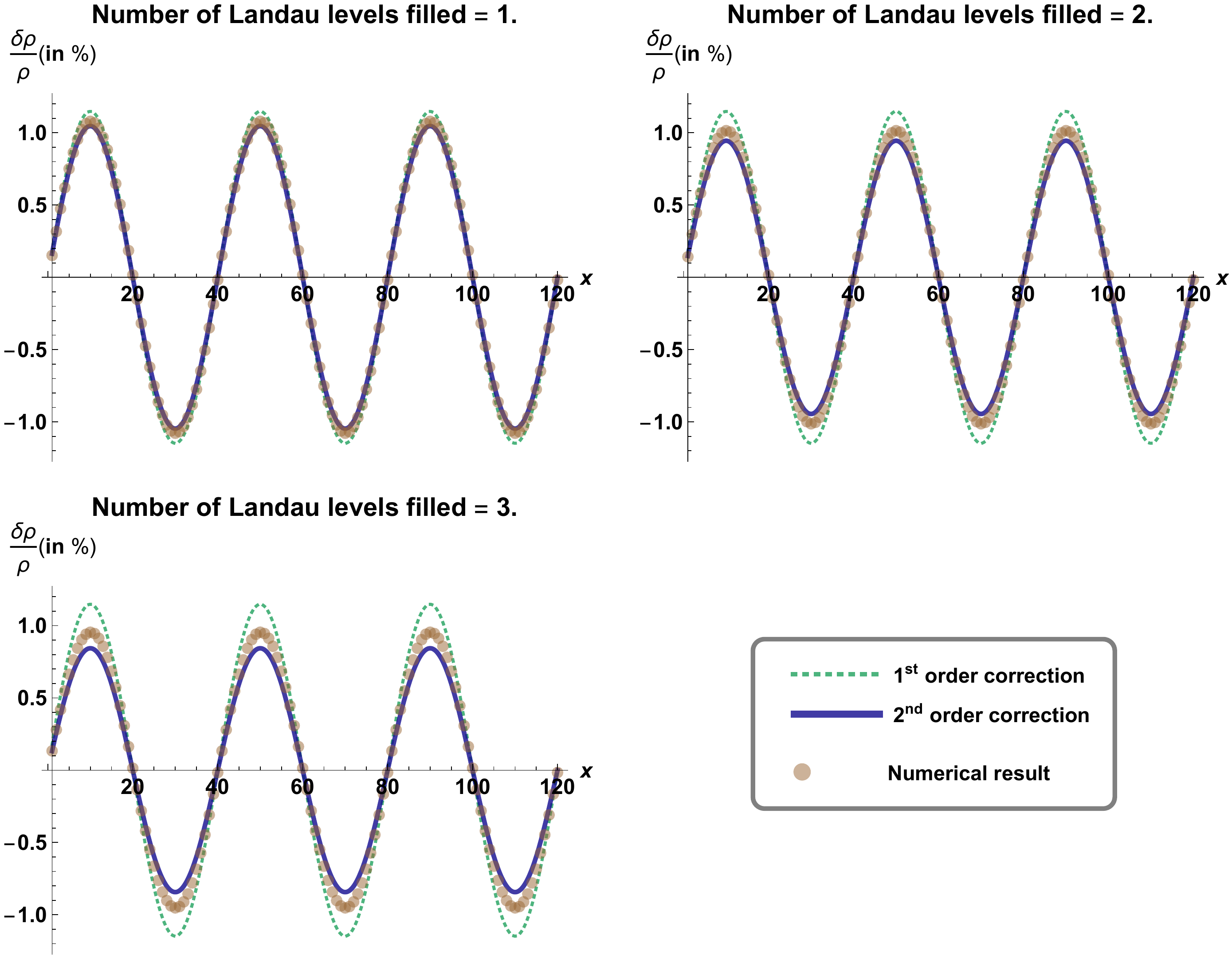}}%inclgphcs[trim=lcm bcm rcm tcm, clip=true, angle=-90]
\caption{The spatial variation of local fractional charge density modulation in a nonuniform (sinusoidal) electric field. The Landau levels are modeled by the lowest bands in a Hofstadter model on a square lattice. The nonuniform electric field is generated by a sinusoidal background potential, which is small compared to the inter-Landau-level energy gap so that the system is in the linear response regime. The brown circles correspond to the local charge density values obtained via numerical diagonalization. The dashed green curve is the response obtained correct to the second derivative in the electric field, Eq.~\eqref{eq-chargeresponsepartial}, corresponding to the nonuniform polarization induced by cyclotron orbit shifts. The thick blue curve corresponds to Eqs.~\eqref{eq-chargeresponse} and \eqref{eq-chargeresponse1d}, correct to the third order in the derivatives of the electric field. For these plots, $V_{0}/\e_{c} = 0.05$, $k\ell = 0.49$.}
\label{fig-rhoplotgrid}
\end{center}
\end{figure}
%%%%%%%%%%%%%%%%%%%%%%%end%%%%%%%%%%%%%%%%%%%%%%

In the presence of a potential that varies only in the $x$-direction, only the $y$-component of the local current density is nonzero. Its value for a filled Landau level with index $n$ is (using Eq.~\eqref{eq-totalcurrentdensity}):
%%%%%%%%%%%%%%%%%%%%%%begin%%%%%%%%%%%%%%%%%%%%%%
\begin{align}%\label{eq-}
j_{y}^{(n)}(\mbx) &= k V_{0} \cos k x \le(1 - \frac{3 (k\ell)^{2}}{2}\le(n + \frac{1}{2}\ri) + \ldots\ri).\nn
\end{align}
%%%%%%%%%%%%%%%%%%%%%%%end%%%%%%%%%%%%%%%%%%%%%%
A convenient observable is the total current denisty when the first $N$ Landau levels are completely filled. It is obtained by summing the preceding expression over $n=0, 1, \ldots, N-1$:
%%%%%%%%%%%%%%%%%%%%%%begin%%%%%%%%%%%%%%%%%%%%%%
\begin{align}\label{eq-totalcurrentdensity1d}
J_{y}^{(N)}(\mbx) &= N k V_{0} \cos k x \le(1 - \frac{3 N }{4}(k\ell)^{2} + \ldots\ri).
\end{align}
%%%%%%%%%%%%%%%%%%%%%%%end%%%%%%%%%%%%%%%%%%%%%%
We have used $\sum_{n=0}^{N-1}\vx_{n} = N^{2}/2$. The expression outside the brackets is the expected current profile for uniform Hall conductance. In Figure~\ref{fig-curplotgrid} we have shown the numerical verification of this result for the lowest three Landau levels.

Finally, our prediction (Eq.~\eqref{eq-chargeresponse}) for the change in local charge density, $\d\rr^{(n)}(x) = \rr^{(n)}(x) - \rr_{0}$, becomes:
%%%%%%%%%%%%%%%%%%%%%%begin%%%%%%%%%%%%%%%%%%%%%%
\begin{align}%\label{eq-}
\frac{\d\rr^{(n)}(x)}{\rr_{0}} &= (k\ell)^{2}\le(1 - \frac{3(k\ell)^{2}}{4}\le(n + \frac{1}{2}\ri) + \ldots \ri) \frac{V_{0} \sin k x}{\e_{c}}.\nn
\end{align}
%%%%%%%%%%%%%%%%%%%%%%%end%%%%%%%%%%%%%%%%%%%%%%
We have used the field-free cyclotron gap, $\e_{c} = 4\p/q$, to scale quantities with dimensions of energy. The second term in the bracket corresponds to the fourth-order derivative of the potential. Previously, we provided a conjecture for the value of its coefficient by connecting it to the topological gravitational response of quantum Hall states: as expected, it is found to be half the gravitational coupling constant. Again, a convenient observable is the fractional change in the total charge density, $\rr_{\text{tot}}^{(N)}(x)$, when the first $N$ Landau levels are full:
%%%%%%%%%%%%%%%%%%%%%%begin%%%%%%%%%%%%%%%%%%%%%%
\begin{align}\label{eq-chargeresponse1d}
\frac{\d\rr^{(N)}_{\text{tot}}(x)}{\rr_{0}^{\text{tot,}N}} &= (k\ell)^{2}\le(1 - \frac{3N}{8} (k\ell)^{2}+ \ldots \ri) \frac{V_{0} \sin k x}{\e_{c}}.
\end{align}
%%%%%%%%%%%%%%%%%%%%%%%end%%%%%%%%%%%%%%%%%%%%%%
In Figure~\ref{fig-rhoplotgrid}, we have shown the numerical verification of this relation for the lowest three Landau levels.

\section{Discussion}\label{sec-discussion}

In this work we have introduced a quantum Hilbert space representation based on gauge-invariant variables (GIV), defined in Eq.~\eqref{eq-GIV}, to describe Schr\"{o}dinger quantum mechanics of two-dimensional charged particles (electrons) in the presence of a uniform perpendicular magnetic field. We have included a background electrostatic potential, which gives rise to a non-uniform electric field, and ignored interactions. The case with interactions has been discussed elsewhere \cite{2018-chen-zl}. The GIV representation is gauge-invariant, exploits the unique geometry of quantum mechanics in the presence of a magnetic field (captured by the GIV commutation relations, Eq.~\eqref{eq-GIVcommutation0}), and provides a natural quantum representation which builds on the classical cyclotron orbit picture and the classical motion of the orbit in an external electric field.

Using the GIV representation we have derived a geometric picture of the response of cyclotron orbits to a non-uniform electric field, as summarized in Fig.~\ref{fig-cyclotronorbit}. To recapitulate, the orbits get sheared, are shifted from their original position, and drift in a direction perpendicular to the shift. These modifications are characterized by an effective shearing metric, $g$, and a vector field, $\mbs{\Dd}$, which controls both orbit drift and location shift; $g$ and $\mbs{\Dd}$ are defined in Eqs.~\eqref{eq-inducedmetric} and \eqref{eq-inducedshift}, respectively.

We have combined this geometric picture and the Wigner quasiprobability formalism to calculate the linear local responses to the nonuniform electric field, as gradient expansions to the second order in derivatives of the electric field. Specifically, we calculated the the local cyclotron orbit energy (Eq.~\eqref{eq-cyclotronenergy}), the local current density (Eq.~\eqref{eq-totalcurrentdensity}), and the local charge density (Eq.~\eqref{eq-chargeresponse}).

These calculations provide mechanistic insights as to why the gravitational coupling constant (defined as $\k$ in Eq.~\eqref{eq-tgr}) appears in the current response to a nonuniform electric field \cite{2012-hoyos-fk,2009-read-fk}. Motivated by our calculation of the local current density response, we were led to the conjecture that the current contribution from the shearing of the cyclotron orbit is the same as the previously-obtained current contribution involving the gravitational coupling constant \cite{2012-hoyos-fk}. Following this, we pursued a stronger conjecture -- that the metric induced by non-uniform electric fields acts upon the quantum Hall state in the same way as a bona fide real-space metric with a Gaussian curvature given by Eq.~\eqref{eq-inducedcurvature} -- in the context of charge density response to a non-uniform electric field. We found that the gravitational coupling constant appears in the local charge density response and enters the electric field gradient expansion for charge response at the third order, Eq.~\eqref{eq-chargeresponseconjecture}. It will be illuminating to see if this conjecture continues to hold in other kinds of quantum Hall states, say, the Laughlin states, for which no such calculation exists. If the conjecture is found to hold generally, it will have implications for improving the extended universal effective theory of quantum Hall states, which in present form \cite{2012-hoyos-fk} does not predict charge response to the third order in the derivative of the electric field.

\begin{acknowledgments}
Author contributions: This research was conceived of and designed by RRB. YK, GJ and RRB performed calculations and wrote the paper. We acknowledge useful discussions with the late Leo Kadanoff, Nick Read, Mike Stone, Paul Wiegmann and Srividya Iyer-Biswas. This research was supported by Purdue University Startup Funds and the Purdue Research Foundation.
\end{acknowledgments}

\end{document}